\documentstyle[12pt]{article}
\advance \textheight by \topskip
\makeatletter
\def\eqnarray{\stepcounter{equation}\let\@currentlabel=\theequation
\global\@eqnswtrue
\global\@eqcnt\z@\tabskip\@centering\let\\=\@eqncr
$$\halign to \displaywidth\bgroup\@eqnsel\hskip\@centering
  $\displaystyle\tabskip\z@{##}$&\global\@eqcnt\@ne
  \hfil$\displaystyle{\hbox{}##\hbox{}}$\hfil
  &\global\@eqcnt\tw@ $\displaystyle\tabskip\z@
  {##}$\hfil\tabskip\@centering&\llap{##}\tabskip\z@\cr}
\@addtoreset{equation}{section}
  \def\theequation{\thesection.\arabic{equation}}
\makeatother
\def\Cal#1{{\cal #1}}
\def\u#1{\raisebox{-0.08ex}{$\displaystyle
  \mathop{u}^{\scriptscriptstyle #1}$}{}}
\def\J#1{\raisebox{0.21ex}{$\displaystyle
  \mathop{J}^{\scriptscriptstyle #1}$}{}}

\begin{document}

\begin{titlepage}
\hbox to \hsize{\hfil hep-th/9306142}
\hbox to \hsize{\hfil IHEP 93--69}
\hbox to \hsize{\hfil May, 1993}
\vfill
\large \bf
\begin{center}
THE CANONICAL SYMMETRY AND \\
HAMILTONIAN FORMALISM. \\
II. HAMILTONIAN OPERATORS
\end{center}
\vskip 1cm
\normalsize
\begin{center}
{\bf A. N. Leznov and A. V. Razumov\footnote{E--mail:
razumov@mx.ihep.su}}\\
{\small Institute for High Energy Physics, 142284 Protvino, Moscow Region,
Russia}
\end{center}
\vskip 2.cm
\begin{abstract}
\noindent
It is shown how the canonical symmetry is used to look for the hierarchy
of the Hamiltonian operators relevant to the system under consideration.
It appears that only the invariance condition can be used to solve the
problem.
\end{abstract}
\vfill
\end{titlepage}

\section{Introduction}

It has been recently observed \cite{Lez91a,Lez92} that most of nonlinear
integrable equations, not only two--dimensional ones, possess a special
discrete symmetry, that appears very useful in many respects.  In
Ref.~\cite{LRa92} we demonstrated that this symmetry considered as a
transformation of the phase space of the system is a canonical
transformation. This fact was one of the reasons to call the considered
transformations the canonical symmetry.  In the next paper \cite{LRa93} we
considered the behaviour of the densities of local conservation laws under
the canonical symmetry for the case of the nonlinear Schr\"odinger
equation. These densities appeared to change under the action of
the canonical symmetry by the total space derivatives. We have noted in
\cite{LRa93} that this observation does not give us a constructive
method to find the local conservation laws for the system under
consideration.

In the present paper we continue the investigation of the properties of
the canonical symmetry from the Hamiltonian point of view. In particular,
we show how the canonical symmetry can be used to look for different
Hamiltonian operators relevant to the system. It is known that if two
Hamiltonian operators, that can be used to write the considered evolution
equations as the Hamilton equations, form a Hamiltonian pair, then we can
find the whole hierarchy of the involutive conservation laws, starting
from the given one \cite{Olv86}. It appears that the use of the canonical
symmetry leads us to a set of the Hamiltonian operators, any two of which
form a Hamiltonian pair. Thus, the canonical symmetry not only gives a new
characterization of the local conservation laws but also helps us to find
them.

In this paper we use the notations and conventions of our paper
\cite{LRa93}. Thus, we consider the case of two independent variables $t$,
$x$ and $A$ dependent variables $u^a$, $a = 1, \ldots, A$. We denote by
$\Cal A$ the space of the functions of the variables $x$, $u^a$, and the
derivatives of $u^a$ over $x$ up to some finite order. For a general
function of such type we use the following notation
\begin{equation}
f[u] = f(x, u, \u{(1)}, \ldots, \u{(K)}), \label{1.1}
\end{equation}
where for any $k$ $\u{(k)}$ is the set formed by the $k$--th derivatives
$\u{(k)}^a$ of the dependent variables $u^a$ over $x$. We assume also that
$\u{(0)}^a \equiv u^a$. Recall that we call the Hamilton equations
evolution equations, that can be written in the form
\begin{equation}
u^a_t = J^{ab}[u] \frac{\delta h[u]}{\delta u^b}, \label{1.2}
\end{equation}
where $J$ is a Hamiltonian operator \cite{Olv86,LRa92}. In the present
paper we consider Hamiltonian operators that include inverse powers of the
operator of the total derivative over $x$. To deal with such operators we
have to use a generalization of the usual binomial coefficients. This
generalization is reminded of in section 2. In section 3 we give the
necessary facts on integro--differential operators. Section 4 is
devoted to the consideration of the invariance of a matrix
integro--differential operator under the action of a differential
transformation. This condition for the case of a Hamiltonian operator is
equivalent to the requirement of the canonicity of the transformation. In
section 5 we consider a partial case of the second order transformations.
Using as examples the nonlinear Schr\"odinger equation and modified
nonlinear Schr\"odinger equation we find all matrix skew-symmetric
operators which satisfy the invariance condition for the corresponding
canonical symmetry. It appears that they are actually the Hamiltonian
operators, forming the well known hierarchies of Hamiltonian operators for
the nonlinear equations under consideration.

\section{Binomial Coefficients}

Here we define the binomial coefficients $\displaystyle {n \choose k}$ for
arbitrary integers $n$ and $k$, such that $n \ge k$. Recall that for $n, k
\ge 0$ and $n \ge k$ we have
\begin{equation}
{n \choose k} = \frac{n!}{k! (n-k)!}. \label{2.1}
\end{equation}
Let us write the well known relation
\begin{equation}
(1 + x)^n = \sum_{k = 0}^n {n \choose k} x^k. \label{2.2}
\end{equation}
It is convenient to put for $k < 0$
\begin{equation}
{n \choose k} = 0. \label{2.3}
\end{equation}
Hence, we can rewrite Eq.~(\ref{2.2}) in the form
\begin{equation}
(1 + x)^n = \sum_{k = -\infty}^n {n \choose k} x^k. \label{2.4}
\end{equation}
We use this formula as the definition of the binomial coefficient for $n <
0$. In particular, for $n = -1$ we have
\begin{equation}
(1 + x)^{-1} = \sum_{k = -\infty}^{-1} (-1)^{-k-1} x^k, \label{2.5}
\end{equation}
hence, for $k \le -1$
\begin{equation}
{-1 \choose k} = (-1)^{-k-1}. \label{2.6}
\end{equation}

For any $n < 0$ we can write
\begin{equation}
(1 + x)^n = (-1)^{-n-1}\frac{1}{(-n-1)!} \frac{d^{-n-1}}{dx^{-n-1}}
(1+x)^{-1}. \label{2.7}
\end{equation}
Differentiating Eq.~(\ref{2.5}), we get
\begin{equation}
{n \choose k} = (-1)^{-n - k} {-k -1 \choose -n -1}. \label{2.8}
\end{equation}
{}From this relation we have subsequently
\begin{equation}
{n \choose n} = 1, \qquad {n \choose n-1} = n, \qquad {n \choose n-2} =
\frac{n(n -1)}{2!}. \label{2.9}
\end{equation}
It is clear that for any $k > 0$
\begin{equation}
{n \choose n-k} = \frac{n(n - 1)\ldots (n-k+1)}{k!}. \label{2.10}
\end{equation}
Note, that equality (\ref{2.10}) is valid also for $n > 0$.


\section{Integro--Differential Operators}

In this section we shall give necessary facts on integro--differential
operators. A more detailed exposition can be found, for example, in
\cite{Man79}. An integro--differential operator is a formal series of the
form
\begin{equation}
a = \sum_{k=-\infty}^{N(a)} a_k D^k, \label{3.1}
\end{equation}
where $a_k \in {\cal A}$. We suppose that $a_{N(a)} \ne 0$, and say that
the number $N(a)$ is the order of the operator $a$.

The multiplication of integro--differential operators is defined as
follows. The product of two integro--differential operators $a$ and $b$ is
the integro--differential operator
\begin{equation}
c = ab = \sum_{k=-\infty}^{N(c)} c_k D^k, \label{3.2}
\end{equation}
where $N(c) = N(a) + N(b)$, and
\begin{eqnarray}
c_k = (ab)_k &\equiv& \sum_{l=k-N(b)}^{N(a)} \sum_{m=k-N(b)}^l {l \choose
m} a_l D^{l-m} (b_{k-m}) \nonumber \\
&=& \sum_{l=k-N(a)}^{N(b)} \sum_{m=k-N(a)}^l {k-m \choose k-l} a_{k-m}
D^{l-m} (b_l). \label{3.3}
\end{eqnarray}
This relation can be formally obtained by the multiplication of the
corresponding formal series and by the appropriate resummation. The direct
calculation shows that thus defined product is associative.

The transposed operator $a^T$ of the integro--differential operator $a$ is
the operator
\begin{equation}
a^T = \sum_{k=-\infty}^{N(a)} (a^T)_k D^k, \label{3.4}
\end{equation}
where
\begin{equation}
(a^T)_k \equiv \sum_{l=k}^{N(a)} (-1)^l {l \choose k} D^{l-k}(a_l).
\label{3.5}
\end{equation}
In particular, we have
\begin{equation}
D^T = -D. \label{3.6}
\end{equation}

We shall need below the following relation
\begin{eqnarray}
(ab^T)_k &=& \sum_{l=k-N(a)}^{N(b)} \sum_{m=k}^{l+N(a)} (-1)^l {m \choose k}
a_{m-l} D^{m-k}(b_l) \nonumber \\
&=& \sum_{l=k-N(b)}^{N(a)} \sum_{m=k}^{l+N(b)} (-1)^{m-l} {m \choose k}
a_l D^{m-k}(b_{m-l}). \label{3.7}
\end{eqnarray}

An integro--differential operator $a$ is said to be invertible if there
exists an integro--differential operator $b$ such that
\begin{equation}
ba = 1. \label{3.8}
\end{equation}
If it is the case we write $b = a^{-1}$. It is clear that $N(a^{-1}) =
-N(a)$.

Suppose that an operator $a$ is invertible, then from Eq.~(\ref{3.3}) for
$k=0$ we get
\begin{equation}
(a^{-1})_{-N(a)} a_{N(a)} = 1. \label{3.9}
\end{equation}
Hence, the operator $a$ is invertible only if
\begin{equation}
a_{N(a)} \ne 0, \label{3.10}
\end{equation}
and in this case
\begin{equation}
(a^{-1})_{-N(a)} = \frac{1}{a_{N(a)}}. \label{3.11}
\end{equation}
{}From Eq.~(\ref{3.3}) for $k < 0$ we have
\begin{equation}
(a^{-1})_{k-N(a)} a_{N(a)} + \sum_{l=k-N(a)+1}^{-N(a)} \sum_{m=k-N(a)}^l
{l \choose m} (a^{-1})_l D^{l-m} (a_{k-m}) = 0. \label{3.12}
\end{equation}
We can consider Eq.~(\ref{3.12}) as a recursive relation for the
determination of the quantities $(a^{-1})_k$, and conclude from it that
the operator $a$ is invertible if and only if Eq.~(\ref{3.10}) is
satisfied.

The matrix integro--differential operator $A$ is a matrix with the matrix
elements being integro--differential operators. The transposition operation
can be naturally extended to the case of such operators. It
is clear also that a matrix integro--differential operator
\begin{equation}
A = \sum_{k=-\infty}^{N(A)} A_k D^k \label{3.13}
\end{equation}
is invertible if and only if the matrix $A_{N(A)}$ is invertible.

\section{Invariance Condition}

Recall \cite{LRa93} that the invertible differential transformation
\begin{equation}
\widetilde u^a = \varphi^a[u] = \varphi^a(x, u, \u{(1)}, \ldots,
\u{(K)}) \label{4.1}
\end{equation}
is a symmetry of Hamilton equations (\ref{1.2}) if
\begin{equation}
\varphi'[u] J[u] \varphi'^T[u] = J[\varphi[u]] \label{4.2}
\end{equation}
and
\begin{equation}
h[\varphi[u]] - h[u] \in \mbox{Ker }\delta/\delta u. \label{4.3}
\end{equation}
In Eq.~(\ref{4.2}) $\varphi'$ denotes the Fr\'echet derivative of the
differential transformation (\ref{4.1}) \cite{Olv86,LRa93}.

In papers \cite{LRa92,LRa93} we have shown that the canonical symmetry of
a nonlinear system is a symmetry of the corresponding Hamilton equations.
It is known that there are different ways to write integrable equations in
the Hamiltonian form. In this the role of the corresponding Hamiltonians
is played by different conserved quantities. We proved in \cite{LRa93},
for the case of nonlinear Schr\"odinger equation, that the densities of
local conservation laws change under the action of the canonical symmetry
by total space derivatives. In other words, if we consider the density of
a conservation law as the density of the corresponding Hamiltonian, then
we see that it satisfies Eq.~(\ref{4.3}). It is natural to suppose that
the corresponding Hamiltonian operator satisfies Eq.~(\ref{4.2}), which we
call the invariance condition. We shall not demonstrate that directly, but
consider a more general problem: what the Hamiltonian operators are that
satisfy Eq.~(\ref{4.2}) for given differential transformation (\ref{4.1}).

In this paper we consider only the case of two dependent variables $u^a$
and use the notation $q \equiv u^1$, $r \equiv u^2$. In this case the
operator $J$ can be written as
\begin{equation}
J = \left(\begin{array}{cc}
a & b \\
c & d
\end{array}\right). \label{4.4}
\end{equation}
The condition $J = -J^T$ gives
\begin{eqnarray}
&&a^T = -a, \qquad d^T = -d, \label{4.5} \\
&&c^T = -b, \qquad b^T = -c, \label{4.6}
\end{eqnarray}
thus,  we can write
\begin{equation}
J = \left(\begin{array}{cc}
a & b \\
-b^T & d
\end{array}\right). \label{4.7}
\end{equation}

The operator $\varphi'$ in the case under consideration may be
represented as
\begin{equation}
\varphi' = \left( \begin{array}{cc}
\alpha & \beta \\
\gamma & \delta
\end{array} \right), \label{4.8}
\end{equation}
and invariance condition (\ref{4.2}) is equivalent to the relations
\begin{eqnarray}
&\widetilde a& = \alpha a \alpha^T - \beta b^T \alpha^T + \alpha b \beta^T
+ \beta d \beta^T, \label{4.9} \\
&\widetilde b& = \alpha a \gamma^T - \beta b^T \gamma^T + \alpha b
\delta^T + \beta d \delta^T, \label{4.10} \\
&- \widetilde b^T& = \gamma a \alpha^T - \delta b^T \alpha^T + \gamma b
\beta^T + \delta d \beta^T, \label{4.11} \\
&\widetilde d& = \gamma a \gamma^T - \delta b^T \gamma^T + \gamma b
\delta^T + \delta d \delta^T, \label{4.12}
\end{eqnarray}
where $\widetilde a[u] \equiv a[\varphi[u]]$, etc.  Note that
Eq.~(\ref{4.11}) can be obtained from Eq.~(\ref{4.10}) via the
transposition.  Thus, we have only three independent relations. In all the
cases that we consider in this paper $\alpha = 0$ and $\beta$, $\gamma$
are simply functions. Therefore, we can rewrite system
(\ref{4.9})--(\ref{4.12}) in the form
\begin{eqnarray}
&&\widetilde a - \beta d \beta = 0, \label{4.13} \\
&&\delta d + \widetilde b^T \beta^{-1} + \gamma b = 0, \label{4.14} \\
&&\delta \beta^{-1} \tilde b + \gamma b
\delta^T + \gamma a \gamma - \widetilde d = 0. \label{4.15}
\end{eqnarray}
{}From this system we conclude, in particular, that
\begin{equation}
N(b) = N(a) + N(\delta), \qquad N(d) = N(a). \label{4.16}
\end{equation}
Representing $a$ and $d$ as
\begin{equation}
a = \sum_{k = -\infty}^{N(a)} a_k D^k, \qquad d = \sum_{k = -\infty}^{N(d)}
d_k D^k, \label{4.17}
\end{equation}
and using Eq.~(\ref{3.3}), we get from Eq.~(\ref{4.13}) the following
equalities
\begin{equation}
\widetilde a_k - \sum_{m=k}^{N(d)} {m \choose k} \beta D^{m-k} (\beta)\,d_m
= 0, \qquad k \le N(a). \label{4.18}
\end{equation}
In the same way, using Eqs.~(\ref{3.3}) and (\ref{3.5}), we write
Eq.~(\ref{4.14}) in the component form $(k \le N(b))$:
\begin{eqnarray}
\sum_{l=k-N(d)}^{N(\delta)} \sum_{m=k-N(d)}^l && {l \choose m} \delta_l
D^{l-m}(d_{k-m}) \nonumber \\
&&{} + \sum_{m=k}^{N(b)} (-1)^m {m \choose k} D^{m-k} (\beta^{-1}
\widetilde b_m) + \gamma b_k = 0. \label{4.19}
\end{eqnarray}
Using now Eq.~(\ref{3.7}), we get from Eq.~(\ref{4.15})
for $N(a) < k \le N(b)+N(\delta)$ the relations
\begin{eqnarray}
\sum_{l=k-N(b)}^{N(\delta)} && \sum_{m=k-N(b)}^l \left({l \choose m}
\delta_l D^{l-m} (\beta^{-1} \widetilde b_{k-m}) \right. \nonumber \\
&&{} + \left. (-1)^l {m+N(b) \choose k} \gamma D^{m+N(b)-k}(\delta_l)
b_{m+N(b)-l} \right) = 0, \label{4.20}
\end{eqnarray}
while for $k \le N(a)$ we have
\begin{eqnarray}
\sum_{l=k-N(b)}^{N(\delta)} \sum_{m=k-N(b)}^l && \left({l \choose m}
\delta_l D^{l-m} (\beta^{-1} \widetilde b_{k-m}) \right. \nonumber \\
&&{} + \left. (-1)^l {m+N(b) \choose k} \gamma D^{m+N(b)-k}(\delta_l)
b_{m+N(b)-l} \right)
\nonumber \\
&& + \sum_{m=k}^{N(a)} {m \choose k} \gamma D^{m-k}(\gamma) a_m -
\widetilde d_k = 0. \label{4.21}
\end{eqnarray}

\section{Invariance Condition for Second Order Transformations}

\subsection{General Consideration}

In this section we consider invariance condition
(\ref{4.13})--(\ref{4.15}) for the case $N(\delta) = 2$, i. e. for the
case when
\begin{equation}
\delta = \delta_2 D^2 + \delta_1 D + \delta_0. \label{5.1}
\end{equation}

First, suppose that there exists a Hamiltonian operator $\J{(0)}$ of
form (\ref{4.7}) with the matrix elements being functions. In other words,
we suppose that there exists a Hamiltonian operator of the order 0, having
a special form. From Eq.~(\ref{4.16}) it follows that in this case $a = d =
0$. Denote the function, playing the role of the operator $b$, by $g$, and
consider Eqs.~(\ref{4.13})--(\ref{4.15}).  Eq.~(\ref{4.13}) is satisfied
by definition. In the case under consideration Eq.~(\ref{4.14}) takes the
form
\begin{equation}
\widetilde g + \beta \gamma g = 0. \label{5.2}
\end{equation}
Note, that in all known cases the canonical symmetry has the following
property. The function $f[u]$, satisfying the relation
\begin{equation}
f[\varphi[u]] = f[u], \label{5.3}
\end{equation}
is simply a constant. From this property it follows that if there exists a
function $g$, satisfying Eq.~(\ref{5.2}), then this function is unique up
to a constant factor.

Eq.~(\ref{4.15}) in our case is equivalent to
\begin{equation}
\delta \gamma g = \gamma g \delta^T. \label{5.4}
\end{equation}
It is easy to show that the function $g$ satisfies Eq.~(\ref{5.4}) if and
only if it satisfies the relation
\begin{equation}
D(\delta_2) \gamma g - \delta_2 D(\gamma g) - \delta_1 \gamma g = 0,
\label{5.5}
\end{equation}
that can be written as
\begin{equation}
D \left( \frac{\delta_2}{\gamma g} \right) = \frac{\delta_1}{\gamma g}.
\label{5.6}
\end{equation}

Proceed now to the investigation of invariance condition
(\ref{4.13})--(\ref{4.15}), for the case when the order of the operator
$b$ is an arbitrary integer $N$. Denote the corresponding operator by
$\J{(N)}$. It is a priory unknown, that this operator exists. On the
other hand, if it exists then it is defined at least up to an arbitrary
linear combination of the operators $\J{(k)}$ with $k < N$.

Let us suppose that the operator $\J{(N)}$ exists.  From Eq.~(\ref{4.20})
in the case $k = N+2$ we get
\begin{equation}
\widetilde b_N + \beta \gamma b_N = 0. \label{5.7}
\end{equation}
Thus for any $N$ the function $b_N$ is determined up to a constant factor
and is proportional to the function $g$. Thus, taking into account
Eq.~(\ref{5.5}) we see that
\begin{equation}
D(\delta_2) \gamma b_N - \delta_2 D(\gamma b_N) - \delta_1 \gamma b_N = 0.
\label{5.8}
\end{equation}

For $k = N+1$ from Eq.~(\ref{4.20}) we get the following relation
\begin{eqnarray}
\delta_2 (\widetilde b_{N-1} - \beta \gamma b_{N-1}) + ((N&+&2)D(\delta_2) -
\delta_1) \beta \gamma b_N \nonumber \\
&+& \delta_1 \widetilde b_N + 2 \delta_2 \beta D(\beta^{-1} \widetilde
b_N) = 0. \label{5.9}
\end{eqnarray}
Using Eqs.~(\ref{5.7}) and (\ref{5.8}), we can rewrite this equality in the
form
\begin{equation}
\delta_2(\widetilde b_{N-1} + \beta\gamma b_{N-1}) + N \beta\gamma
D(\delta_2) b_N = 0. \label{5.10}
\end{equation}
It is convenient to introduce the functions
\begin{equation}
f_k = b_k/b_N. \label{5.11}
\end{equation}
For the function $f_{N-2}$ from Eq.~(\ref{5.10}) we have
\begin{equation}
\widetilde f_{N-1} - f_{N-1} = N \frac{D(\delta_2)}{\delta_2}. \label{5.12}
\end{equation}
Hence the function $f_{N-1}$, if it exists, is
defined up to an additive constant. The function $b_{N-1}$, in turn, is
defined up to a constant multiplied by the function $b_N$. Recall that the
function $b_N$ is proportional to the function $g$. It is clear, that the
operator $J{(N)}$ is defined up to a linear combination of the operators
$J{(k)}$ with $k < N$. It is the reason of the ambiguity that we have come
across.

After some rather long calculations from Eq.~(\ref{4.20}) for the case $k
= N$ we get the following equality
\begin{eqnarray}
\delta_2(\widetilde b_{N-2} &+& \beta\gamma b_{N-2}) - 2\delta_2
\beta\gamma D\left( \frac{b_{N-1}}{b_N} \right) b_N - N \delta_2
\beta\gamma D\left(\frac{\delta_1}{\delta_2} \right) b_N
\nonumber \\
&+& (N-1) \beta \gamma D(\delta_2) b_{N-1} + \frac{1}{2} N(N-1)
\beta\gamma D^2(\delta_2) b_N = 0, \label{5.13}
\end{eqnarray}
that can be written in terms of the functions $f_k$ as
\begin{eqnarray}
\widetilde f_{N-2} &-& f_{N-2} = {} - 2D(f_{N-1}) -
ND\left(\frac{\delta_1}{\delta_2}\right) \nonumber \\
&+& (N-1) \frac{D(\delta_2)}{\delta_2} f_{N-1} + \frac{1}{2} N(N-1)
\frac{D^2(\delta_2)}{\delta_2}. \label{5.14}
\end{eqnarray}
We see that the function $b_{N-2}$ is defined ambiguously, and the reason
of this ambiguity is the same as for the function $b_{N-1}$ (see above).

Consider now Eq.~(\ref{4.19}). For $k=N$ we get
\begin{equation}
d_{N-2} = ((-1)^N - 1) \frac{\gamma b_N}{\delta_2}. \label{5.15}
\end{equation}
Thus for an odd $N$ we have
\begin{equation}
d_{N-2} = {} - \frac{2\gamma b_N}{\delta_2}, \label{5.16}
\end{equation}
and for an even $N$
\begin{equation}
d_{N-2} = 0. \label{5.17}
\end{equation}
Note, that we have not actually used the fact that the operator $d$ must
be skew--symmetric, but Eq.~(\ref{5.17}) is already in agreement
with it.

We get further
\begin{equation}
d_{N-3} = {} - (N-2)D\left(\frac{\gamma b_N}{\delta_2}\right) =
\frac{N-2}{2} D(d_{N-2}) \label{5.18}
\end{equation}
for an odd $N$, and
\begin{equation}
d_{N-3} = {} - \frac{2 \gamma b_{N-1}}{\delta_2} - \frac{N \delta_1 \gamma
b_N}{\delta_2^2} \label{5.19}
\end{equation}
for an even $N$. Eq.~(\ref{5.18}) is again in agreement with the
skew--symmetricity of the operator $d$.

At last we get
\begin{eqnarray}
d_{N-4} = {} - \frac{2\gamma b_{N-2}}{\delta_2} &+& \frac{2 \delta_1
\gamma b_{N-1}}{\delta_2^2} + (N+1) D\left( \frac{\gamma
b_{N-1}}{\delta_2}\right) \nonumber \\
&+& \frac{2\delta_0 \gamma b_N}{\delta_2^2} - \frac{1}{2} (N^2-3N+4) D^2
\left( \frac{\gamma b_N}{\delta_2} \right) \label{5.20}
\end{eqnarray}
for an odd $N$, and
\begin{equation}
d_{N-4} = \frac{N-3}{2} D(d_{N-3}) \label{5.21}
\end{equation}
for an even $N$.

The functions $a_k$ can be determined from Eq.~(\ref{4.18}).

Let us make here some conclusions from our consideration of the invariance
condition.  The ambiguity in the definition of the operator $\J{(N)}$ can
be described as follows. Let for any $k$ the operator $\J{(k)}$ be some
solution of the invariance condition. A general operator, satisfying the
invariance condition, and having the order $N$, is given by the formula
\begin{equation}
J = \sum_{k = -\infty}^{N} \nu_k \J{(k)},
\end{equation}
where $\nu_k$ are some constants.

If we have two solutions, for example $\J{(0)}$ and $\J{(1)}$, of the
invariance condition, and the operator $\J{(0)}$ is invertible, then the
operator
\begin{equation}
\J{(2)} = \J{(1)} \J{(0)}^{-1} \J{(1)}
\end{equation}
is of the order two and also satisfies the invariance condition.
Analogously, we get the operators, having any positive order. If the
operator $\J{(1)}$ is invertible, then we can construct an operator of
any negative order. For example, the operator of the order $-1$ has the
form
\begin{equation}
\J{(-1)} = \J{(0)} \J{(1)}^{-1} \J{(0)}.
\end{equation}

\subsection{Nonlinear Schr\"odinger Equation}

As a first concrete example, we consider the nonlinear Schr\"odinger
equation. In fact to define the canonical symmetry we should consider the
following complex extension of it \cite{LRa92}:
\begin{equation}
i\dot q + q'' - 2 \epsilon r q^2 = 0, \qquad i \dot r - r'' + 2 \epsilon
q r^2 = 0, \label{6.1}
\end{equation}
where $q$ and $r$ are arbitrary complex functions of the variables $x$ and
$t$, $\epsilon$ is the coupling constant. In Eq.~(\ref{6.1}) and below dot
and prime mean the partial derivative over $t$ and $x$, respectively. The
canonical symmetry for this system has the form \cite{Lez91a,Lez92,SYa91}
\begin{equation}
\widetilde q = \frac{1}{\epsilon r}, \qquad \widetilde r = \epsilon r^2 q -
r'' + \frac{r'^2}{r}. \label{6.2}
\end{equation}
Thus, we have
\begin{eqnarray}
&\beta = - \frac{1}{\epsilon r^2}, \qquad \gamma = \epsilon r^2,&
\label{6.3} \\
&\delta = -D^2 + 2 \frac{r'}{r} D + 2 \epsilon rq - \frac{r'^2}{r^2}.&
\label{6.4}
\end{eqnarray}

Equations (\ref{6.1}) can be written in Hamiltonian form (\ref{1.2}) if we
choose
\begin{equation}
J = \J{(0)} = \left( \begin{array}{rr}
0 & -i  \\
i & 0 \end{array} \right), \label{6.5}
\end{equation}
and
\begin{equation}
h[q,r] = r' q' + \epsilon r^2 q^2. \label{6.6}
\end{equation}
It is not difficult to show that the operator $\J{(0)}$ satisfies the
invariance condition \cite{LRa92,LRa93}. Hence, in this case the function
$g$ is a constant function.

Recall \cite{LRa92,LRa93} that for the conserved quantity $rq$ we have
\begin{equation}
\widetilde r \widetilde q = rq - D\left(\frac{r'}{\epsilon r}\right).
\label{6.7}
\end{equation}

Consider now the case $N=1$. From Eq.~(\ref{5.7}) we get
\begin{equation}
\widetilde b_1 - b_1 = 0, \label{6.8}
\end{equation}
hence, the function $b_1$ is a constant function. It is convenient to
normalize the operator $\J{(1)}$ by the condition
\begin{equation}
b_1 = -1. \label{6.9}
\end{equation}
{}From Eq.~(\ref{5.10}) we have
\begin{equation}
\widetilde b_0 - b_0 = 0. \label{6.10}
\end{equation}
Let us put
\begin{equation}
b_0 = 0. \label{6.11}
\end{equation}
Eq.~(\ref{5.13}) is more nontrivial:
\begin{equation}
\widetilde b_{-1} - b_{-1} + 2 D\left(\frac{r'}{r}\right) = 0.
\label{6.12}
\end{equation}
Comparing this relation with Eq.~(\ref{6.7}), we see that we can choose
\begin{equation}
b_{-1} = 2\epsilon r q. \label{6.13}
\end{equation}

{}From Eqs.~(\ref{5.16}), (\ref{5.18}) and (\ref{5.20}) we find
\begin{equation}
d_{-1} = -2\epsilon r^2, \qquad d_{-2} = 2\epsilon rr', \qquad d_{-3} =
- 2\epsilon r r'', \label{6.14}
\end{equation}
while Eq.~(\ref{4.18}) gives
\begin{equation}
a_{-1} = -2\epsilon q^2, \qquad a_{-2} = 2\epsilon q q', \qquad a_{-3} =
- 2\epsilon q q''. \label{6.15}
\end{equation}

Taking into account the expressions obtained for $b_k$, $d_k$ and $a_k$ we
can suppose that the operator $\J{(1)}$ has the form
\begin{equation}
\J{(1)} = \left( \begin{array}{cc}
-2\epsilon q D^{-1} q & -D + 2\epsilon q D^{-1} r \\
-D + 2 \epsilon r D^{-1} q & -2\epsilon r D^{-1} r
\end{array} \right). \label{6.16}
\end{equation}
A direct calculation shows that the operator, given by Eq.~(\ref{6.16})
satisfies the invariance condition. Thus, we get the well known second
Hamiltonian operator for the nonlinear Schr\"odinger equation
\cite{Mag78,FTa87} as a solution of the invariance condition.

\subsection{Modified Nonlinear Schr\"odinger Equation}

We consider here the following complex extension \cite{Lez91a,Lez92,LRa92}
of the modified nonlinear Schr\"odinger equation \cite{CLL79}
\begin{equation}
i \dot q + q'' - 2i \epsilon (rq) q' = 0, \qquad i\dot r - r'' - 2i \epsilon
(rq) r' = 0. \label{7.1}
\end{equation}
The canonical symmetry for this system is the transformation of the form
\cite{Lez91a,Lez92,LRa92}
\begin{equation}
\widetilde q = \frac{1}{\epsilon r}, \qquad \widetilde r = \epsilon r^2 q + i
\left(r' - \frac{r'' r}{r'}\right). \label{7.2}
\end{equation}
Thus, in this case
\begin{eqnarray}
&\beta = - \frac{1}{\epsilon r^2}, \qquad \gamma = \epsilon r^2,& \label{7.3}
\\
&\delta = -i \frac{r}{r'} D^2 + i \left( 1 + \frac{r''r}{r'^2}
\right) D + 2 \epsilon rq - i \frac{r''}{r'}.&
\label{7.4}
\end{eqnarray}

Equations (\ref{7.1}) can be written in Hamiltonian form (\ref{1.2}) for
the Hamiltonian operator $J$, given by Eq.~(\ref{6.5}) and
\begin{equation}
h[q,r] =   r' q' + \frac{i\epsilon}{2} (r^2 q q' - q^2 r r').
\label{7.5}
\end{equation}
It is not difficult to show that the operator $\J{(0)}$ satisfies again
the corresponding invariance condition \cite{LRa92}, and the function $g$
is again a constant function.

Write now the formulae, describing the behaviour of the densities of the
lowest conservation laws under the action of the canonical symmetry:
\begin{eqnarray}
&\widetilde r \widetilde q = r q + \frac{i}{\epsilon} D \left(\ln
\frac{r}{r'}\right),& \label{7.6} \\
&-i \widetilde r \widetilde q' = -i r q' + D \left(i r q +
\frac{r'}{\epsilon r}\right).& \label{7.7}
\end{eqnarray}

Proceed now to the construction of the operator $\J{(1)}$. The relation,
determining the function $b_1$, is again (\ref{6.8}), and we can put
\begin{equation}
b_1 = -1. \label{7.8}
\end{equation}
{}From Eq.~(\ref{5.12}) we find
\begin{equation}
\widetilde f_0 - f_0 - D\left(\ln \frac{r}{r'}\right) = 0. \label{7.9}
\end{equation}
Taking into account Eq.~(\ref{7.6}), we see that we can put
\begin{equation}
b_0 = i\epsilon r q. \label{7.10}
\end{equation}
Eq.~(\ref{5.14}) takes in our case the form
\begin{equation}
\widetilde f_{-1} - f_{-1} - \epsilon D\left(2i rq + \frac{r'}{\epsilon r} +
\frac{r''}{\epsilon r'} \right) = 0. \label{7.11}
\end{equation}
Using Eq.~(\ref{7.7}), we find the following expression for $b_{-1}$:
\begin{equation}
b_{-1} = i\epsilon (r q' - r' q). \label{7.12}
\end{equation}

As in the case of the nonlinear Schr\"odinger equation we obtain
\begin{eqnarray}
&d_{-1} = 2i\epsilon r r', \quad d_{-2} = -i\epsilon(r'^2 + r r''), \quad
d_{-3} = i\epsilon (r r''' + r' r'');& \label{7.13} \\
&a_{-1} = - 2i\epsilon q q', \quad a_{-2} = i\epsilon(q'^2 + q q''),
\quad a_{-3} = - i\epsilon (q q''' + q' q'').& \label{7.14}
\end{eqnarray}

Thus, we can try as the operator $\J{(1)}$ the operator of form
(\ref{4.7}) with
\begin{eqnarray}
a &=& - i\epsilon (q D^{-1} q' + q' D^{-1} q), \label{7.15} \\
b &=& -D + i\epsilon rq + i \epsilon (r D^{-1} q' - r' D^{-1} q),
\label{7.16} \\
d &=& i\epsilon (r D^{-1} r' + r' D^{-1} r). \label{7.17}
\end{eqnarray}
A direct check shows that this operator satisfies the invariance
condition.

\section{Conclusion}

In this paper we have shown how the canonical symmetry can be used for the
construction of Hamiltonian operators relevant to the system under
consideration. Note, that we have used only the invariance condition, and
discovered that the matrix skew--symmetric operators, satisfying this
condition, are the required Hamiltonian operators. These operators define
the Poisson brackets, satisfying Jacobi identity, and any two of them form
a Hamiltonian pair. It is clear that for an arbitrary differential
transformation this will not be the case. An open problem here is to
formulate conditions on the transformation that provide such results.
Apparently this problem is directly connected to the problem we stated in
\cite{LRa93}, that was: what are the conditions providing the involutivity
of the quasi--invariants of a differential transformation?

\end{document}